\pdfoutput=1
\documentclass[12pt,a4paper]{article}

\usepackage{ifthen} 
\usepackage{placeins}
\newboolean{pdflatex}
\setboolean{pdflatex}{true} 

\newboolean{articletitles}
\setboolean{articletitles}{true} 

\newboolean{uprightparticles}
\setboolean{uprightparticles}{false} 

\def\paperauthors{LHCb collaboration} 
\def\paperasciititle{Modification of Upsilon production in pO and OO collisions at LHCb} 
\def\papertitle{Modification of $\Upsilonres$ production in $p$O and OO collisions at \lhcb} 
\def\paperkeywords{{High Energy Physics}, {LHCb}} 
\def\papercopyright{\the\year\ CERN for the benefit of the LHCb collaboration} 
\def\paperlicence{CC BY 4.0 licence}
\def\paperlicenceurl{https://creativecommons.org/licenses/by/4.0/}

\newif\ifEnableSectionTOCLinks
\EnableSectionTOCLinksfalse 

\usepackage[top=1in, bottom=1.25in, left=1in, right=1in]{geometry}

\usepackage{microtype}
\usepackage{lineno}  
\usepackage{xspace} 
\usepackage{caption} 

\usepackage{graphicx}  
\usepackage{color}
\usepackage{colortbl}
\graphicspath{{./figs/}} 

\usepackage{amsmath} 
\usepackage{amssymb}
\usepackage{amsfonts}
\usepackage{upgreek} 

\newcommand*\patchAmsMathEnvironmentForLineno[1]{%
\expandafter\let\csname old#1\expandafter\endcsname\csname #1\endcsname
\expandafter\let\csname oldend#1\expandafter\endcsname\csname
end#1\endcsname
 \renewenvironment{#1}%
   {\linenomath\csname old#1\endcsname}%
   {\csname oldend#1\endcsname\endlinenomath}%
}
\newcommand*\patchBothAmsMathEnvironmentsForLineno[1]{%
  \patchAmsMathEnvironmentForLineno{#1}%
  \patchAmsMathEnvironmentForLineno{#1*}%
}
\AtBeginDocument{%
\patchBothAmsMathEnvironmentsForLineno{equation}%
\patchBothAmsMathEnvironmentsForLineno{align}%
\patchBothAmsMathEnvironmentsForLineno{flalign}%
\patchBothAmsMathEnvironmentsForLineno{alignat}%
\patchBothAmsMathEnvironmentsForLineno{gather}%
\patchBothAmsMathEnvironmentsForLineno{multline}%
\patchBothAmsMathEnvironmentsForLineno{eqnarray}%
}

\usepackage[pdftex,
            pdfauthor={\paperauthors},
            pdftitle={\paperasciititle},
            pdfkeywords={\paperkeywords}]{hyperref}
\usepackage{hyperxmp}
\hypersetup{
    pdfcopyright={Copyright (C) \papercopyright},
    pdflicenseurl={\paperlicenceurl}
}

\usepackage[colorinlistoftodos,textsize=scriptsize]{todonotes}

\usepackage[bottom,flushmargin,hang,multiple]{footmisc}

\usepackage[all]{hypcap} 

\usepackage{xspace} 
\usepackage{upgreek}

\def\lhcb   {\mbox{LHCb}\xspace}

\def\MagUp {\mbox{\em Mag\kern -0.05em Up}\xspace}

\ifthenelse{\boolean{uprightparticles}}%
{

 \def\Pmu         {\ensuremath{\upmu}\xspace}

 \def\Ppsi        {\ensuremath{\uppsi}\xspace}

 \def\PDelta      {\ensuremath{\Delta}\xspace}                 
 \def\PXi         {\ensuremath{\Xi}\xspace}                 
 \def\PLambda     {\ensuremath{\Lambda}\xspace}                 
 \def\PSigma      {\ensuremath{\Sigma}\xspace}                 
 \def\POmega      {\ensuremath{\Omega}\xspace}                 
 \def\PUpsilon    {\ensuremath{\Upsilon}\xspace}
 \let\oldPi\Pi
 \def\PPi         {\ensuremath{\oldPi}\xspace}

 \def\PB      {\ensuremath{\mathrm{B}}\xspace}                 
 \def\PD      {\ensuremath{\mathrm{D}}\xspace}                 
 \def\PJ      {\ensuremath{\mathrm{J}}\xspace}                 
 \def\PK      {\ensuremath{\mathrm{K}}\xspace}                 
 \def\Ps      {\ensuremath{\mathrm{s}}\xspace}

 \def\thebaroffset{0.0em}
}
{

 \def\Pmu         {\ensuremath{\mu}\xspace}

 \def\Ppsi        {\ensuremath{\psi}\xspace}                 
                  
 \mathchardef\PDelta="7101
 \mathchardef\PXi="7104
 \mathchardef\PLambda="7103
 \mathchardef\PSigma="7106
 \mathchardef\POmega="710A
 \mathchardef\PUpsilon="7107
 \mathchardef\PPi="7105
 \def\PB      {\ensuremath{B}\xspace}                 
 \def\PD      {\ensuremath{D}\xspace}                 
 \def\PJ      {\ensuremath{J}\xspace}                 
 \def\PK      {\ensuremath{K}\xspace}                 
 \def\Ps      {\ensuremath{s}\xspace}

 \def\thebaroffset{0.18em}
}
\newcommand{\offsetoverline}[2][\thebaroffset]{\kern #1\overline{\kern -#1 #2}}%

\makeatletter
\ifcase \@ptsize \relax
  \newcommand{\miniscule}{\@setfontsize\miniscule{4}{5}}
\or
  \newcommand{\miniscule}{\@setfontsize\miniscule{5}{6}}
\or
  \newcommand{\miniscule}{\@setfontsize\miniscule{5}{6}}
\fi
\makeatother

\DeclareRobustCommand{\optbar}[1]{\shortstack{{\miniscule (\rule[.5ex]{1.25em}{.18mm})}
  \\ [-.7ex] $#1$}}

\def\mup        {{\ensuremath{\Pmu^+}}\xspace}
\def\mun        {{\ensuremath{\Pmu^-}}\xspace} 

\def\squark    {{\ensuremath{\Ps}}\xspace}

\def\KorKbar {\kern \thebaroffset\optbar{\kern -\thebaroffset \PK}{}\xspace}

\def\D       {{\ensuremath{\PD}}\xspace}

\def\DorDbar {\kern \thebaroffset\optbar{\kern -\thebaroffset \PD}\xspace}

\def\Dp      {{\ensuremath{\D^+}}\xspace}
\def\Dm      {{\ensuremath{\D^-}}\xspace}

\def\DpDm    {\ensuremath{\Dp {\kern -0.16em \Dm}}\xspace}

\def\B       {{\ensuremath{\PB}}\xspace}

\def\BorBbar {\kern \thebaroffset\optbar{\kern -\thebaroffset \PB}\xspace}

\def\Bd      {{\ensuremath{\B^0}}\xspace}

\def\BdorBdbar {\kern \thebaroffset\optbar{\kern -\thebaroffset \Bd}\xspace}

\def\Bs      {{\ensuremath{\B^0_\squark}}\xspace}

\def\BsorBsbar {\kern \thebaroffset\optbar{\kern -\thebaroffset \Bs}\xspace}

\def\jpsi     {{\ensuremath{{\PJ\mskip -3mu/\mskip -2mu\Ppsi}}}\xspace}

\def\Upsilonres  {{\ensuremath{\PUpsilon}}\xspace}
\def\Y#1S{\ensuremath{\PUpsilon{(#1S)}}\xspace}
\def\OneS  {{\Y1S}\xspace}
\def\TwoS  {{\Y2S}\xspace}
\def\ThreeS{{\Y3S}\xspace}

\def\LorLbar     {\kern \thebaroffset\optbar{\kern -\thebaroffset \PLambda}\xspace}

\def\BF         {{\ensuremath{\mathcal{B}}}\xspace}
\def\BR         {\BF}

\newcommand{\decay}[2]{\ensuremath{\mathinner{#1\!\to #2}}\xspace}

\def\to                 {\ensuremath{\rightarrow}\xspace}

\def\AT#1     {\ensuremath{A_{\mathrm{T}}^{#1}}\xspace}           

\def\C#1      {\ensuremath{\mathcal{C}_{#1}}\xspace}                       
\def\Cp#1     {\ensuremath{\mathcal{C}_{#1}^{'}}\xspace}                    
\def\Ceff#1   {\ensuremath{\mathcal{C}_{#1}^{\mathrm{(eff)}}}\xspace}        
\def\Cpeff#1  {\ensuremath{\mathcal{C}_{#1}^{'\mathrm{(eff)}}}\xspace}       
\def\Ope#1    {\ensuremath{\mathcal{O}_{#1}}\xspace}                       
\def\Opep#1   {\ensuremath{\mathcal{O}_{#1}^{'}}\xspace}                    

\newcommand{\aunit}[1]{\ensuremath{\text{\,#1}}}       

\newcommand{\tev}{\aunit{Te\kern -0.1em V}\xspace}
\newcommand{\gev}{\aunit{Ge\kern -0.1em V}\xspace}
\newcommand{\mev}{\aunit{Me\kern -0.1em V}\xspace}
\newcommand{\kev}{\aunit{ke\kern -0.1em V}\xspace}
\newcommand{\ev}{\aunit{e\kern -0.1em V}\xspace}
 
\newcommand{\mevc}{\ensuremath{\aunit{Me\kern -0.1em V\!/}c}\xspace}
\newcommand{\gevc}{\ensuremath{\aunit{Ge\kern -0.1em V\!/}c}\xspace}
\newcommand{\mevcc}{\ensuremath{\aunit{Me\kern -0.1em V\!/}c^2}\xspace}
\newcommand{\gevcc}{\ensuremath{\aunit{Ge\kern -0.1em V\!/}c^2}\xspace}

\def\nb {\aunit{nb}\xspace}
\def\invnb {\ensuremath{\nb^{-1}}\xspace}
\def\pb {\aunit{pb}\xspace}
\def\invpb {\ensuremath{\pb^{-1}}\xspace}

\def\gsim{{~\raise.15em\hbox{$>$}\kern-.85em
          \lower.35em\hbox{$\sim$}~}\xspace}
\def\lsim{{~\raise.15em\hbox{$<$}\kern-.85em
          \lower.35em\hbox{$\sim$}~}\xspace}

\def\sqs   {\ensuremath{\protect\sqrt{s}}\xspace}
\def\sqsnn {\ensuremath{\protect\sqrt{s_{\scriptscriptstyle\text{NN}}}}\xspace}
\def\pt         {\ensuremath{p_{\mathrm{T}}}\xspace}

\def\evtgen     {\mbox{\textsc{EvtGen}}\xspace}

\def\geant      {\mbox{\textsc{Geant4}}\xspace}

\def\photos     {\mbox{\textsc{Photos}}\xspace}

\def\pythia     {\mbox{\textsc{Pythia}}\xspace}

\def\tell1  {TELL1\xspace}
\def\ukl1   {UKL1\xspace}

\newcommand{\lhcborcid}[1]{\href{https://orcid.org/#1}{\hspace*{0.1em}\raisebox{-0.45ex}{\includegraphics[width=1em]{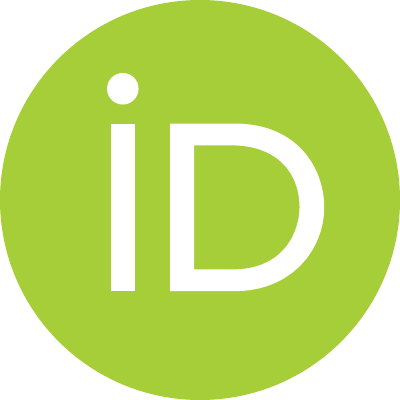}}}}

\hypersetup{
  colorlinks   = true, 
  urlcolor     = blue, 
  linkcolor    = blue, 
  citecolor    = red   
}

\ifEnableSectionTOCLinks
    \usepackage[explicit]{titlesec} 
    
    \let\oldcontentsline\contentsline
    \renewcommand

    \titleformat{\section}{\normalfont\Large\bf}{\hyperlink{tocsection.\thesection}{{\thesection} \parbox[t]{\dimexpr\textwidth-1pc}{#1}}}{1pc}{}

    \titleformat{\subsection}{\normalfont\bf}{\hyperlink{tocsubsection.\thesubsection}{{\thesubsection} \parbox[t]{\dimexpr\textwidth-1pc}{#1}}}{1pc}{}

    \titleformat{name=\section,numberless}[display]{}{}{0pt}{\normalfont\Huge\bfseries #1}
\fi

\usepackage{cite} 
\usepackage{LHCb/mciteplus}

\def\nS  {{\Y nS}\xspace}
\def \epsacc {\epsilon^{\textrm{acc}}}
\def \epstrig{\epsilon^{\textrm{trig}}}
\def \epsrec{\epsilon^{\textrm{rec}}}
\def \epsPID{\epsilon^{\textrm{PID}}}
\def \Ncoll{\ensuremath{\textrm{N}_{\textrm{coll}}}}

\usepackage{longtable} 

\begin{document}

\renewcommand{\thefootnote}{\fnsymbol{footnote}}
\setcounter{footnote}{1}


\begin{titlepage}
\pagenumbering{roman}

\vspace*{-1.5cm}
\centerline{\large EUROPEAN ORGANIZATION FOR NUCLEAR RESEARCH (CERN)}
\vspace*{1.5cm}
\noindent
\begin{tabular*}{\linewidth}{lc@{\extracolsep{\fill}}r@{\extracolsep{0pt}}}
\ifthenelse{\boolean{pdflatex}}
{\vspace*{-1.5cm}\mbox{\!\!\!\includegraphics[width=.14\textwidth]{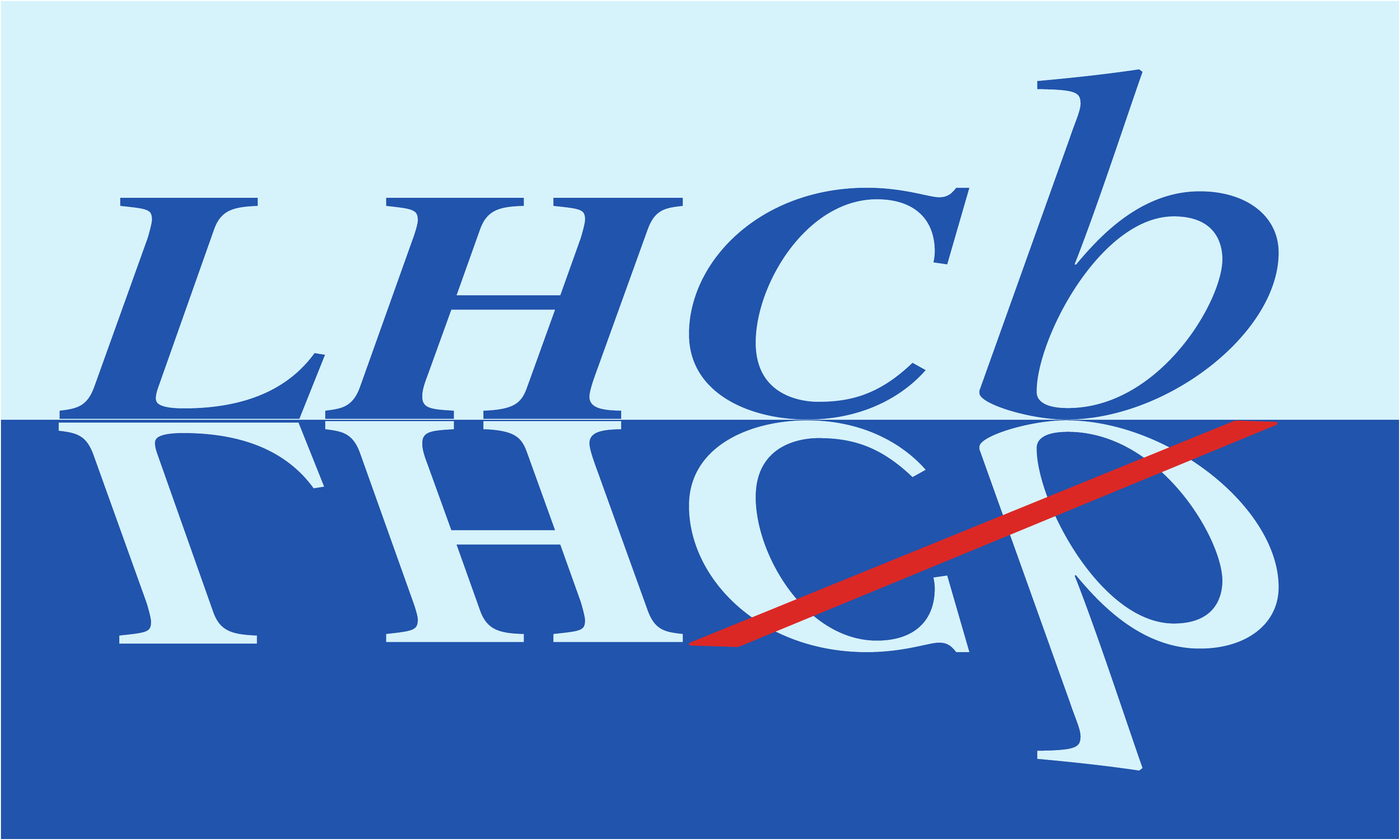}} & &}%
{\vspace*{-1.2cm}\mbox{\!\!\!\includegraphics[width=.12\textwidth]{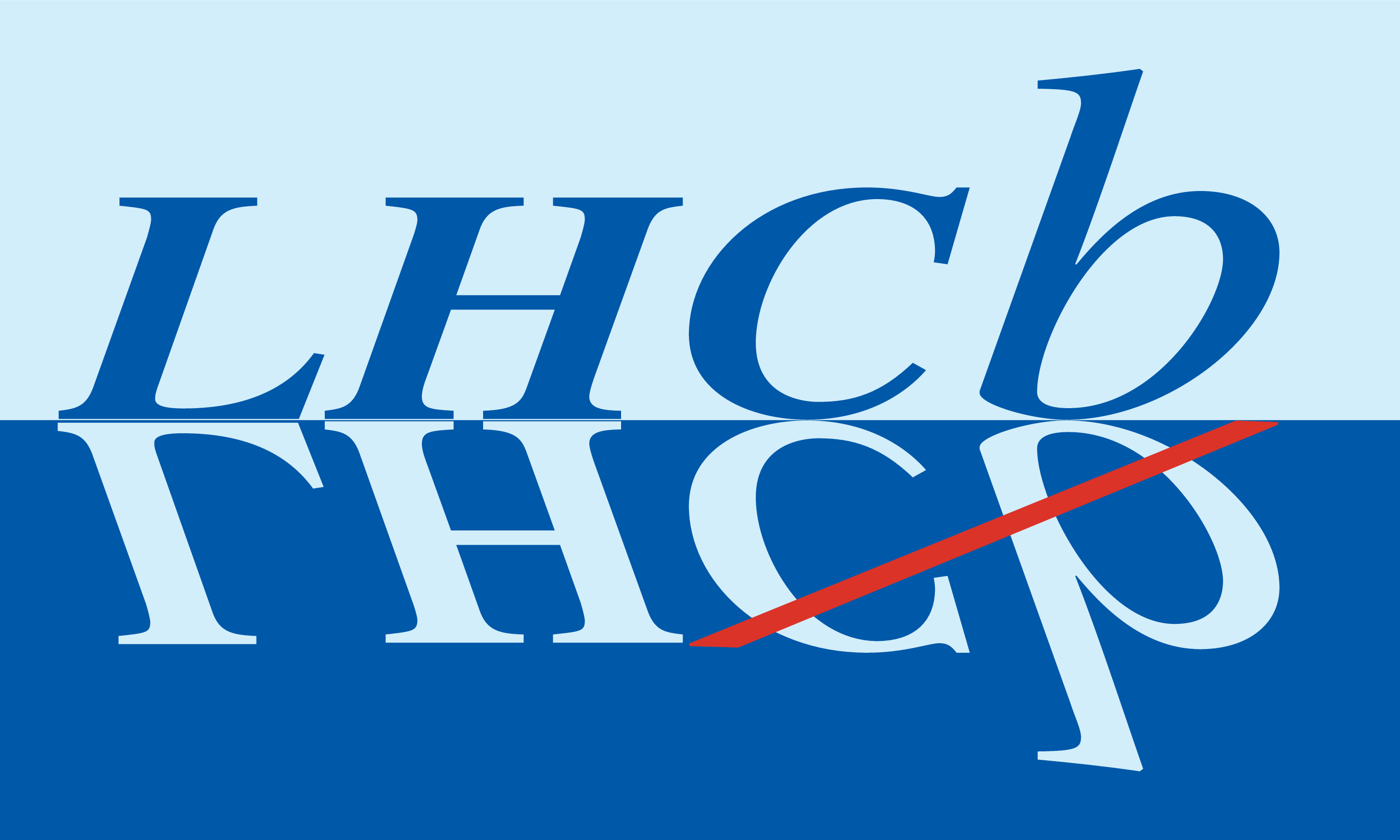}} & &}%
\\
 & & CERN-EP-2026-207 \\  
 & & LHCb-PAPER-2026-027 \\  
 & & \today \\ 
 & & \\
\end{tabular*}

\vspace*{4.0cm}

{\normalfont\bfseries\boldmath\huge
\begin{center}
  \papertitle 
\end{center}
}

\vspace*{2.0cm}

\begin{center}
\paperauthors\footnote{Authors are listed at the end of this Letter.}
\end{center}

\vspace{\fill}

\begin{abstract}
  \noindent
  The production rates of $\mathit{\Upsilon}(2S)$ and $\mathit{\Upsilon}(3S)$ mesons relative to that of the $\mathit{\Upsilon}(1S)$ state are measured in $pp$, $p$O, and OO collisions by the LHCb collaboration.  The ratios  measured in $pp$ data are consistent with previous LHCb measurements at different center-of-mass energies.  Only slight relative suppression of the $\mathit{\Upsilon}(2S)$ and $\mathit{\Upsilon}(3S)$ states is found in $p$O collisions, while in OO collisions the $\mathit{\Upsilon}(2S)$ is suppressed by a factor of $\sim2$, with evidence for suppression of the $\mathit{\Upsilon}(3S)$.  The significant suppression in OO data, compared to the small effect in $p$O data, shows the emergence of additional suppression mechanisms in the relatively small OO collision system.  Models incorporating quark-gluon plasma formation in OO collisions successfully describe the data.  Implications for the interplay between cold nuclear matter effects and color screening in a deconfined quark-gluon plasma are discussed.
\end{abstract}

\vspace*{2.0cm}

\begin{center}
  Submitted to
  Phys.~Rev.~Lett. 
 
\end{center}

\vspace{\fill}

{\footnotesize 
\centerline{\copyright~\papercopyright. \href{\paperlicenceurl}{\paperlicence}.}}
\vspace*{2mm}

\end{titlepage}


\newpage
\setcounter{page}{2}
\mbox{~}
%
%
%
%


\renewcommand{\thefootnote}{\arabic{footnote}}
\setcounter{footnote}{0}


\cleardoublepage


\pagestyle{plain} 
\setcounter{page}{1}
\pagenumbering{arabic}


Collisions of heavy nuclei at ultrarelativistic energies produce an energy density sufficient for nuclear matter to undergo a phase transition to a quark-gluon plasma (QGP), a deconfined state of strongly interacting matter in which quarks and gluons are not confined within hadrons \cite{PHENIX:2004vcz,STAR:2005gfr,BRAHMS:2004adc,PHOBOS:2004zne,Busza:2018rrf,ALICE:2022wpn,CMS:2024krd}. In such a medium, the binding potential between heavy quark–antiquark pairs is screened by exposed color charges, leading to a suppression of quarkonia formation. Excited states, with relatively weak binding energies, are particularly sensitive to these effects. Quarkonia suppression has long been proposed as a signature of QGP formation \cite{Matsui:1986dk,Karsch:1987pv,Kharzeev:1994pz,Gunion:1996qc,Brambilla:2008cx}.

As such, measurements of quarkonia in collisions of nuclei form a cornerstone of experimental high-temperature quantum chromodynamics (QCD).  The \OneS, \TwoS, and \ThreeS mesons, which are bound states of heavy $b$ and $\overline{b}$ quarks, are of special interest.   The three states are experimentally accessible through their decays to $\mup\mun$ pairs, and their different binding energies (about $1.1 \gev$, $0.5 \gev$, and $0.2 \gev$ for the 1S, 2S, and 3S resonances, respectively) induce different sensitivities to disruption in the QGP \cite{Satz:2005hx}. The suppression of higher bottomonia states which feed down to the \nS states also affects the observed yields \cite{LHCb-PAPER-2014-031}.  Measurements from the Relativistic Heavy Ion Collider (RHIC) using gold-gold collisions at a center-of-mass energy per nucleon of $\sqsnn = 200 \gev$ showed suppression of the inclusive \nS yields relative to $pp$ collisions \cite{STAR:2013kwk,PHENIX:2014tbe,STAR:2022rpk}.  Similar levels of suppression have been observed in lead-lead collisions at the Large Hadron Collider (LHC), where the most precise measurements have shown a clear hierarchy: \ThreeS production is more suppressed than \TwoS, which in turn is more suppressed than \OneS production \cite{CMS:2012gvv, ALICE:2014wnc, CMS:2017ycw, CMS:2018zza, ALICE:2020wwx, ATLAS:2022exb,CMS:2023lfu}.  This sequential suppression is typically explained in terms of color screening in the QGP, which more strongly affects the relatively weakly bound $2S$ and $3S$ states \cite{Strickland:2011aa, Mocsy:2013syh, Du:2017qkv, Armesto:2026fit}.  

However, the full interpretation of quarkonia suppression in heavy-ion collisions must also account for cold nuclear matter (CNM) effects, which can alter quarkonia production rates without requiring the formation of QGP \cite{Vogt:2010aa,Arleo:2025oos}.  Modifications of parton distributions in nuclei \cite{Kusina:2017gkz,Duwentaster:2022kpv}, energy loss in nuclear matter \cite{Arleo:2012rs,Arleo:2012hn,Arleo:2013zua}, and late-stage breakup through interactions with other produced particles \cite{CapellaComovers,Capella:2000zp,Ferreiro_comovers, CGCcomover} can all contribute to quarkonia suppression, possibly in competition with a short-lived plasma phase \cite{Dinh:2019ajl}.  The first measurements of \nS production with protons on nuclear targets by the E772 experiment showed a suppression that increased with the target mass number A \cite{Alde:1991sw}.  Measurements in deuterium-gold and proton-lead  collisions at RHIC and the LHC, where QGP is not expected to dominate, also show suppression of \Upsilonres production, which affects the $2S$ and $3S$ states more strongly than the $1S$ \cite{STAR:2013kwk, PHENIX:2012xws, LHCb-PAPER-2014-015, ATLAS:2017prf, LHCb-PAPER-2018-035, ALICE:2019qie, CMS:2022wfi}. It is currently not well understood where the crossover lies between the dominance of CNM effects in the small proton-nucleus collision systems versus that of QGP effects in larger collision systems.

One way to study the interplay of these effects is to vary the collision system energy density, which can be done by comparing different colliding-beam species.  Since both CNM effects and QGP properties depend on the geometry and size of the colliding nuclei \cite{Sharma:2012dy}, comparisons across collision systems provide constraints on the relative contributions of the underlying suppression mechanisms.  To this end, the LHC delivered the first proton-oxygen~($p$O), oxygen-oxygen~(OO), and neon-neon~(NeNe) collisions during Run-3 in 2025, which provide a testing ground for constraining CNM effects and searching for the onset of QGP formation in small collision systems.  Thus far, strong collective effects have been observed in OO collisions, which are well described by hydrodynamic models \cite{ATLAS:2025nnt,ALICE:2025luc,CMS:2025tga,CMS:2026sai}. Particle production in OO is suppressed with respect to $pp$ collisions \cite{CMS:2025bta,CMS:2026qef,ATLAS:2026fos,ALICE:2026zck}, which may be due to the energy loss of partons as they cross the QGP.  The production of prompt charmed mesons in OO and NeNe collisions is inconsistent with the effects of modified nuclear parton distributions alone \cite{LHCb-PAPER-2026-008}, requiring additional effects from charm quark energy loss in QGP for a successful description of the data \cite{Ke:2022gkq}.  Suppression of \nS states at midrapidity has been found in OO and NeNe collisions \cite{CMS:2026qbb}, but the interplay with potential CNM effects  has not been explored. 

This Letter describes measurements of the production of the bottomonia states \TwoS and \ThreeS, relative to that of the \OneS state, in $pp$, $p$O, and OO collisions at the \lhcb experiment.  The $pp$ and OO data were recorded at a center-of-mass energy of $\sqsnn = 5.36\tev$ per nucleon, with integrated luminosities of about 230\invpb and 5.5\invnb, respectively.  The $p$O data were collected at $\sqsnn = 9.62\tev$ per nucleon and correspond to a luminosity of about 33\invnb.  The \Upsilonres states are reconstructed through the decays \decay{\nS}{\mup\mun}. These data enable studies of the baseline bottomonia production in vacuum in $pp$, the influence of CNM effects in $p$O, and the potential emergence of QGP in OO collisions.

The \lhcb Run~3 detector is a single-arm forward spectrometer covering the pseudorapidity range $2 < \eta < 5$, described in detail in Ref.~\cite{LHCb-DP-2022-002}.  Muon candidates are identified as tracks that originate from the primary vertex and have momentum $p>3\gev$ with a transverse component with respect to the beam axis $\pt>650\mev$.  Natural units $\hbar$=\textit{c}=1 are used throughout the Letter. To reject hadrons, tracks are required to penetrate the muon system and not be consistent with the pion hypothesis~\cite{LHCb-TDR-014}.  The rapidity $y$ of the resulting dimuon candidate is required to be within $2<y<4.5$ for the $pp$ and OO data, and within $1.6<y<4.1$ for the $p$O data, due to the rapidity shift caused by the asymmetric beam energies. 

Simulation is required to model the effects of the detector acceptance and the imposed selection requirements.  In the simulation, $pp$ collisions are generated using \pythia \cite{Sjostrand:2007gs} with a specific \lhcb configuration~\cite{LHCb-PROC-2010-056}, while $p$O and OO collisions are generated using EPOS~\cite{Werner:2023zvo}.  Decays of unstable particles are described by \evtgen~\cite{Lange:2001uf}, in which final-state radiation is generated using \photos~\cite{davidson2015photos}.  The interaction of the generated particles with the detector, and its response, are implemented using the \geant toolkit~\cite{Allison:2006ve, *Agostinelli:2002hh} as described in Ref.~\cite{LHCb-PROC-2011-006}.

\begin{figure}[!ht]
  \centering
  \includegraphics[width=1.0\linewidth]{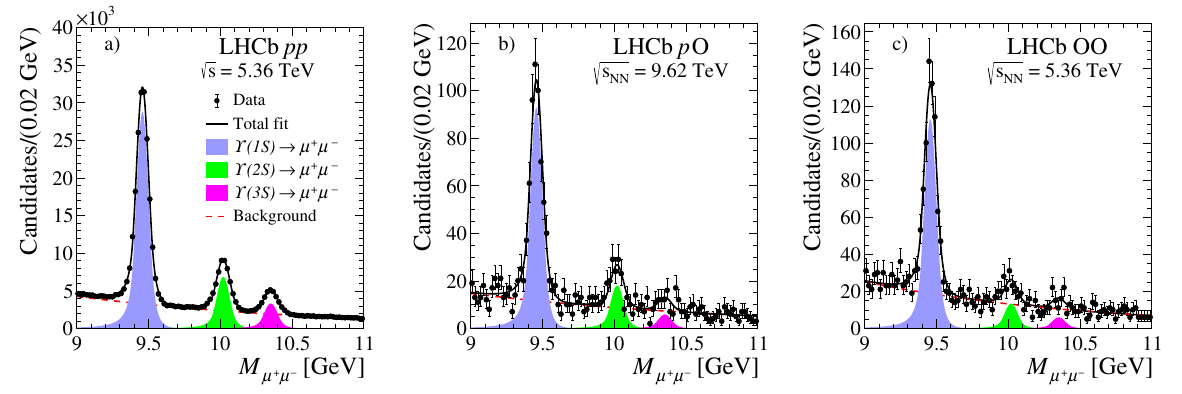}
  \caption{Dimuon invariant mass spectra from a) $pp$ collisions at $\sqs = 5.36\tev$, b) $p$O collisions at $\sqsnn = 9.62\tev$, and c) OO collisions at $\sqsnn = 5.36\tev$. Projections of a fit modeling the \nS resonances and the background are also shown.} 
  \label{fig:Fig1}
\end{figure}

The selected dimuon candidates invariant mass $M_{\mup \mun}$ spectra from the $pp$, $p$O, and OO data are shown in Fig.~\ref{fig:Fig1}. The yields of the \nS states $N_{\nS}$ are measured through an extended binned maximum-likelihood fit to the invariant mass spectra.  The signal distributions are parameterized as Crystal Ball functions with both low- and high-mass power-law tails, where the tail parameters are fixed to values obtained from simulation. The background is a mixture of uncorrelated $\mup\mun$ pairs and pairs produced by the Drell-Yan process and semileptonic decays of charm and bottom hadrons, and is modeled by an exponential function.  The invariant mass spectra from each collision system are fit simultaneously, and the signal functions are constrained to have the same shape across all systems, while the signal yields and background parameters are allowed to vary. The mass differences between the \nS states are fixed to the known values from the PDG~\cite{PDG2024}, and the widths $\sigma(nS)$ are constrained to scale linearly with the fitted width of the \OneS with mass $M(nS)$ as $\sigma(nS) = \sigma(1S) \times M(nS)/M(1S)$.  The resulting fit is shown along with the data in Fig.~\ref{fig:Fig1}.

The yields extracted from the fit must be corrected for detector efficiencies.   The ratio of production cross sections in the dimuon decay channel is

\begin{equation}
\frac{\sigma_{\nS}}{\sigma_{\OneS}} \times
\frac{\BR[\decay{\nS}{\mup \mun}]}{\BR[\decay{\OneS}{\mup \mun}]}
=
\frac{N_{\nS}}{N_{\OneS}}
\times
\frac{\epsacc_{\OneS}}{\epsacc_{\nS}}
\times
\frac{\epstrig_{\OneS}}{\epstrig_{\nS}}
\times
\frac{\epsrec_{\OneS}}{\epsrec_{\nS}}
\times
\frac{\epsPID_{\OneS}}{\epsPID_{\nS}},
\end{equation}
where $\epsacc_{\OneS}/\epsacc_{\nS}$ is the ratio of the acceptance of the LHCb spectrometer for the ${\nS}$ decay products, $\epstrig_{\OneS}/\epstrig_{\nS}$ is the ratio of efficiencies for triggering on the dimuons, $\epsrec_{\OneS}/\epsrec_{\nS}$ is the ratio of efficiencies for reconstructing the $\mup \mun$ pair, and $\epsPID_{\OneS}/\epsPID_{\nS}$ is the ratio of muon identification efficiencies.  The ratio of acceptance corrections is found to be consistent with unity in simulation. Since the $\pt$ distribution of the $\nS$ states has never been measured in $p$O and OO collisions, and the kinematics of the parent $\nS$ states affect the $\mup \mun$ kinematics, the generated $\nS$ distributions are weighted to match measurements from $p$Pb and PbPb collisions \cite{LHCb-PAPER-2018-035,CMS:2018zza}.  This varies the ratios of acceptance corrections by less than 1\%, which is rounded up to 1\% and taken as a systematic uncertainty.  A trigger that requires at least one track reconstructed in the silicon-pixel vertex detector is used when recording $p$O and OO data, giving a ratio of trigger efficiencies of one.  The $pp$ data use a high-level trigger in software that requires a dimuon candidate with high invariant mass.  These trigger efficiencies are measured to be consistent with unity, using the techniques discussed in Refs. \cite{LHCb-PUB-2014-039,Albrecht:2025une}.  The statistical uncertainty on the data used to determine the ratio of trigger efficiencies is taken as a systematic uncertainty on the $\nS$ ratios, which is 3\% for the $\TwoS/\OneS$ ratio and 6\% for the $\ThreeS/\OneS$ ratio.  The ratios of reconstruction efficiencies are determined to be consistent with unity using simulation, and compared to reconstruction efficiencies measured using tag-and-probe methods from the data \cite{LHCb-DP-2013-002}.  The difference of 1\% between these methods is taken as a systematic uncertainty on the ratios of cross sections.  The muon identification efficiencies are measured using known muons from \decay{\jpsi}{\mup\mun} decays, and have a ratio that is consistent with unity with an uncertainty of 1\% that is due to the finite size of the calibration samples.  In all cases, the resulting product of the four ratios of efficiencies is consistent with unity.  The largest uncertainty on the $pp$ data is the systematic uncertainty on the ratio of trigger efficiencies, while the $p$O and OO data are dominated by statistical uncertainties.

 \begin{figure}
  \centering
  \includegraphics[width=0.6\linewidth]{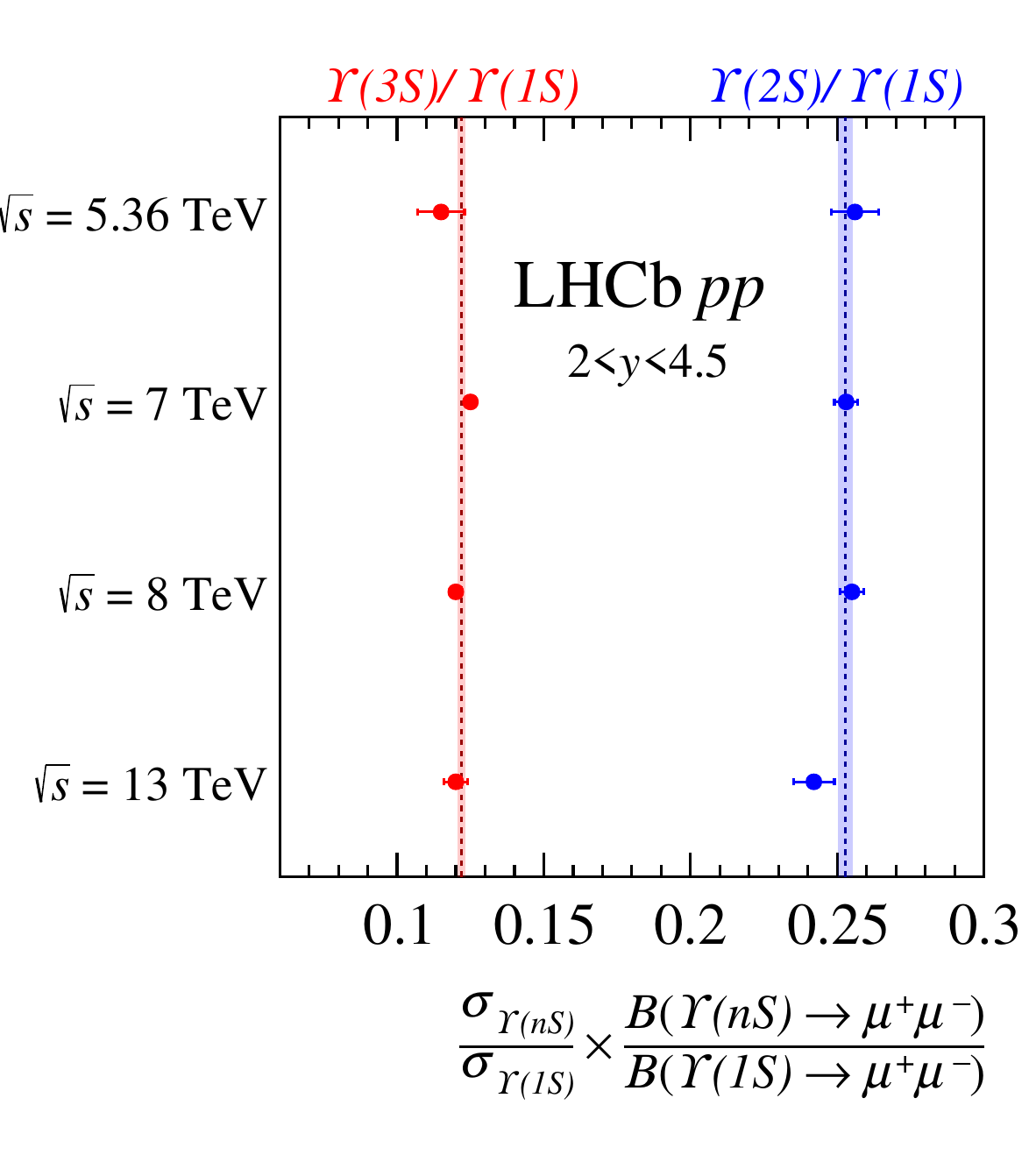}
  \caption{Ratios of the \ThreeS/\OneS and \TwoS/\OneS cross sections times branching fraction to dimuons in $pp$ collisions, from the measurement described here at $\sqs = 5.36 \tev$, and previous LHCb measurements  \cite{LHCb-PAPER-2015-045, LHCb-PAPER-2018-002}.  The vertical lines show the weighted average.} 
  \label{fig:Fig2}
\end{figure}

The resulting ratio of cross sections times branching fractions is shown in Fig.~\ref{fig:Fig2} for the $pp$ data, where statistical and systematic uncertainties are added in quadrature and shown as an error bar.  These data and previous measurements from the LHCb experiment at higher collision energies~\cite{LHCb-PAPER-2015-045, LHCb-PAPER-2018-002} are found to agree well.  Models based on nonrelativistic QCD that successfully describe total $\nS$ cross sections expect little or no $\sqs$ dependence in these ratios, since the long-distance matrix elements governing quarkonium hadronization are universal \cite{Bodwin:1994jh,Brambilla:2010cs,Lansberg:2019adr}.  The weighted average of the ratios measured across different energies are shown as lines, with values 
\begin{equation}
\left\langle\frac{\sigma_{\TwoS}}{\sigma_{\OneS}} \times
\frac{\BR\!\left[\TwoS \to \mu^+ \mu^-\right]}
     {\BR\!\left[\OneS \to \mu^+ \mu^-\right]} \right\rangle
= 0.2527 \pm 0.0025
\end{equation}

\noindent and 

\begin{equation}
\left\langle\frac{\sigma_{\ThreeS}}{\sigma_{\OneS}} \times
\frac{\BR\!\left[\ThreeS \to \mu^+ \mu^-\right]}
     {\BR\!\left[\OneS \to \mu^+ \mu^-\right]}\right\rangle
= 0.1220 \pm 0.0013.
\end{equation}

\begin{figure}[!ht]
  \centering
  \includegraphics[width=0.6\linewidth]{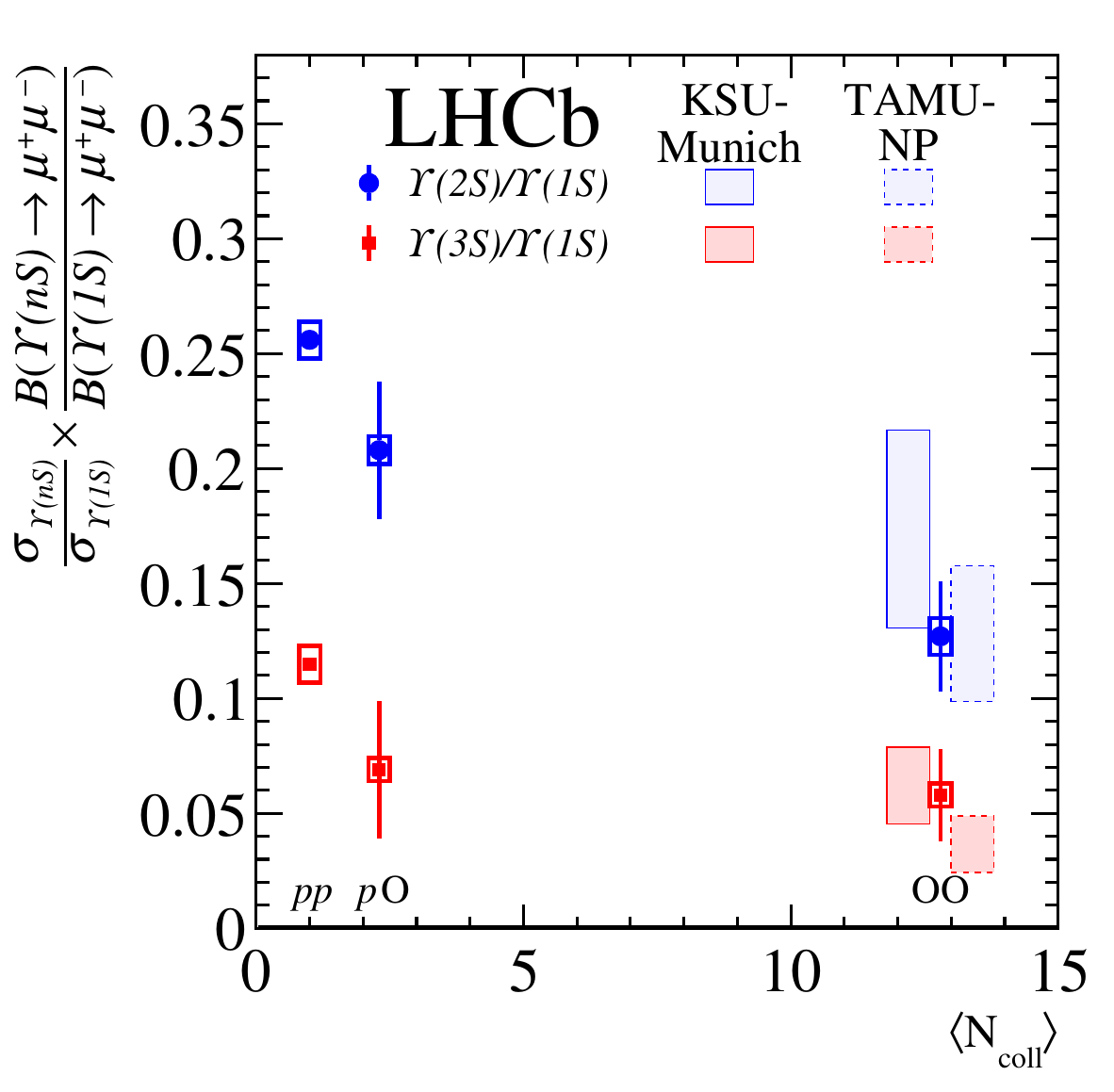}
  \caption{Ratios of the \ThreeS/\OneS and \TwoS/\OneS production cross sections in the dimuon decay channel from $pp$ and OO collisions at $\sqsnn = 5.36\tev$ and $p$O collisions at $\mbox{\sqsnn = 9.62\tev}$. The error bars (boxes) on the data represent the statistical (systematic) uncertainties.  Two model calculations for OO collisions are also shown offset on the horizontal axis for clarity, where the height of the boxes represent theoretical uncertainties.}   
  \label{fig:Fig3}
\end{figure}

The ratios from $pp$, $p$O, and OO collisions are shown together in Fig.~\ref{fig:Fig3} as a function of the average number of binary nucleon-nucleon collisions in each system $\langle\Ncoll\rangle$ obtained from a Glauber model of the oxygen nucleus\cite{Loizides:2025ule}.  On this plot, statistical uncertainties are shown as error bars while systematic uncertainties are represented by boxes.  Numerical values of the ratios are given in the Supplemental Material.  Initial-state effects, such as modifications of the nuclear parton distribution function, are expected to largely cancel in these ratios.  In $p$O collisions, both the $\ThreeS/\OneS$ and $\TwoS/\OneS$ ratios show a slight suppression with respect to $pp$, which could indicate the emergence of CNM effects that preferentially suppress the \TwoS and \ThreeS states more than the \OneS.  However, the uncertainties on the $p$O data preclude drawing firm conclusions.  The OO data show strong relative suppression of the \TwoS state which differs from the $pp$ data by a factor of two.  This suppression goes beyond the relatively small effect seen in $p$O collisions, indicating that additional suppression mechanisms beyond CNM effects come into play in the larger OO collision system.  This could be due to the emergence of QGP in OO collisions. The \ThreeS/\OneS ratio in OO collisions is also suppressed relative to $pp$ by a factor of about two, but is consistent with the suppression seen in $p$O collisions within uncertainties.  

Predictions from two different state-of-the-art models are shown for comparison with the OO data.  The KSU-Munich model (displayed as a solid box in Fig.~\ref{fig:Fig3}) is an open quantum-system model that includes CNM effects from the modification of nuclear parton distribution functions and coherent energy loss of partons crossing the nucleus, and final-state effects from bottomonia interactions with the QGP \cite{Strickland:2024oat,Omar:2021kra,Brambilla:2022ynh}. The in-medium bottomonia suppression is calculated by solving the Lindblad equation within the potential nonrelativistic QCD effective field theory. The TAMU-NP model (shown as a dashed box) is a semiclassical transport model that includes the same CNM effects with suppression and regeneration of bottomonia in the QGP \cite{Thapa:2025jua,Wu:2025lcj}. Here, suppression of the bottomonia states is calculated by solving a kinetic rate equation, with reaction rates that are constrained by lattice QCD.  Both models agree well with the measured  \TwoS/\OneS and \ThreeS/\OneS ratios in OO collisions.

\begin{figure}[!ht]
  \centering
  \includegraphics[width=\linewidth]{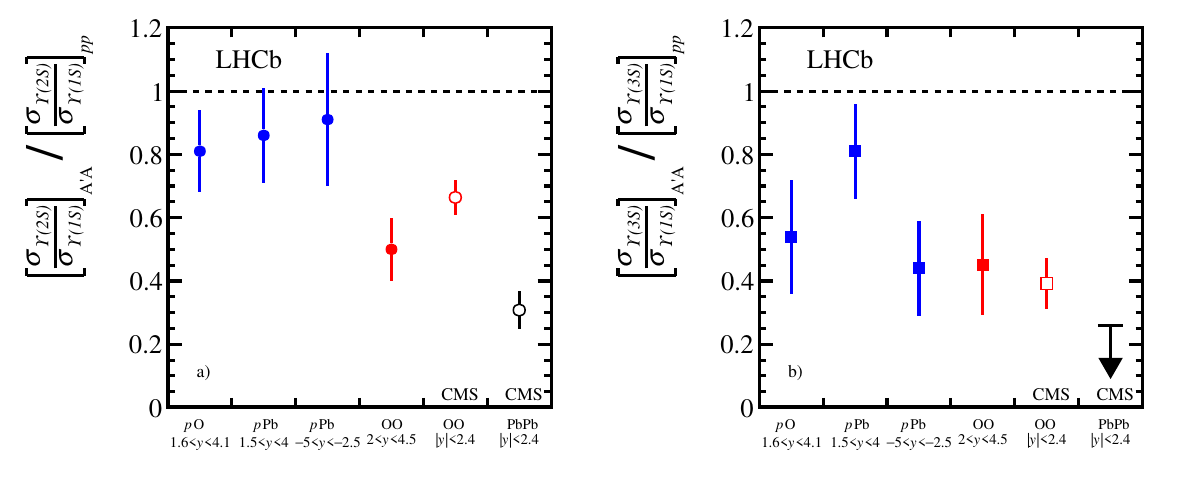}
  \caption{Double ratios of the a) \TwoS/\OneS and b) \ThreeS/\OneS cross sections from (blue) $p$O and $p$Pb collisions at LHCb~\cite{LHCb-PAPER-2018-035}, (red) OO collisions at LHCb and CMS \cite{CMS:2026qbb}, and (black) PbPb collisions at CMS\cite{CMS:2017ycw}.} 
  \label{fig:Fig4}
\end{figure}

A direct comparison between the data from ion and $pp$ collisions is accomplished by measuring the double ratios 
\begin{equation}
 \left[\frac{\sigma_{\nS}}{\sigma_{\OneS}}\right]_{\mathrm{A}'\mathrm{A}} \Big/  \left[\frac{\sigma_{\nS}}{\sigma_{\OneS}}\right]_{pp},    
\end{equation}
    
\noindent where A$'$A refers to the collision system with ions. The double ratios are shown in Fig.~\ref{fig:Fig4}, where the statistical and systematic uncertainties have been added in quadrature and are represented with the error bar.  Numerical values are given in the Supplemental Material. The \TwoS/\OneS double ratios in $p$A collisions show little suppression, indicating that CNM effects are relatively small.  The OO data show significant suppression, and differs from one by $\sim4.8$ standard deviations.   The suppression level observed in OO is consistent with the measurement made at midrapidity by the CMS experiment~\cite{CMS:2026qbb}.  Data from PbPb collisions shows a larger relative suppression of the \TwoS state than is observed in OO collisions~\cite{CMS:2017ycw}, potentially due to more significant QGP effects in the larger collision system. In contrast, the \ThreeS/\OneS double ratio shows suppression across all systems, which is likely due to an interplay of cold nuclear matter and QGP effects.  The \ThreeS state is especially sensitive to these effects due to its smaller binding energy. In OO collisions, the \ThreeS/\OneS double ratio differs from one by $\sim3.5$ standard deviations, providing evidence for suppression of the \ThreeS state in OO collisions.  This is also consistent with CMS measurements at midrapidity, indicating that the suppression effects have little or no rapidity dependence.  Data from PbPb collisions is represented by an arrow showing the upper limit at the 95\% confidence level~\cite{CMS:2017ycw}.

In summary, the LHCb collaboration has measured the ratios of \TwoS and \ThreeS cross sections to the \OneS cross section in $pp$, $p$O, and OO collisions. The \ThreeS/\OneS and \TwoS/\OneS ratios in $pp$ collisions are consistent across the \sqs range measured by LHCb. A strong suppression of the relative rate of \TwoS production is observed in OO collisions, similar to what is seen in PbPb collisions.  This level of suppression is not seen in $p$O collisions, indicating that the observed reduction is not due to CNM effects alone.  The relative production of \ThreeS mesons is suppressed in both $p$O and OO collisions, which could be due to an interplay of CNM and QGP effects on this relatively weakly bound state.  The relative suppression of the \TwoS and \ThreeS mesons in OO collisions is well described by models of that incorporate a deconfined QGP phase. Taken together, these data provide new evidence that is consistent with quark-gluon plasma formation in OO collisions.

\FloatBarrier

\section*{Acknowledgements}
%
%
\noindent We express our gratitude to our colleagues in the CERN
accelerator departments for the excellent performance of the LHC. We
thank the technical and administrative staff at the LHCb
institutes.
We acknowledge support from CERN and from the national agencies:
ARC (Australia);
CAPES, CNPq, FAPERJ and FINEP (Brazil); 
MOST and NSFC (China); 
CNRS/IN2P3 and CEA (France);  
BMFTR, DFG and MPG (Germany);
NKFIH (Hungary);              
INFN (Italy); 
NWO (Netherlands); 
MNiSW and NCN (Poland); 
MEC/IFA (Romania); 
MICIU and AEI (Spain);
SNSF and SER (Switzerland); 
NASU (Ukraine); 
STFC (United Kingdom); 
DOE NP and NSF (USA).
We acknowledge the computing resources that are provided by ARDC (Australia), 
CBPF (Brazil),
CERN, 
IHEP and LZU (China),
IN2P3 (France), 
KIT and DESY (Germany), 
INFN (Italy), 
SURF (Netherlands),
Polish WLCG (Poland),
IFIN-HH (Romania), 
PIC (Spain), CSCS (Switzerland), 
GridPP (United Kingdom),
and NSF (USA).  
We are indebted to the communities behind the multiple open-source
software packages on which we depend.
Individual groups or members have received support from
RTP (Australia), 
FWO Odysseus grant G0ASD25N (Belgium), 
Key Research Program of Frontier Sciences of CAS, CAS PIFI, CAS CCEPP (China); 
Minciencias (Colombia);
EPLANET, Marie Sk\l{}odowska-Curie Actions, ERC and NextGenerationEU (European Union);
A*MIDEX, ANR, IPhU and Labex P2IO, and R\'{e}gion Auvergne-Rh\^{o}ne-Alpes (France);
Alexander-von-Humboldt Foundation (Germany);
ICSC (Italy); 
Severo Ochoa and Mar\'ia de Maeztu Units of Excellence, GVA, XuntaGal, GENCAT, InTalent-Inditex and Prog.~Atracci\'on Talento CM (Spain);
the Leverhulme Trust, the Royal Society and UKRI (United Kingdom).




\section{Supplemental material for LHCb-PAPER-2026-027}
\label{sec:Supplementary-App}
Numerical values of the results are reported in this section. Table~\ref{tab:Tab1} lists the \ThreeS/\OneS and \TwoS/\OneS production cross-section ratios in the dimuon decay channel for \textit{pp}, \textit{p}O and OO systems; Table~\ref{tab:Tab2} shows the values of the double ratios. 

\begin{table}[h]
   \caption{Ratios of the \ThreeS/\OneS and \TwoS/\OneS production cross sections in the dimuon decay channel, as shown in Fig.~\ref{fig:Fig3}.  The first uncertainty is statistical and the second is systematic. }
\begin{center}\begin{tabular}{|c|c|c|}
    \hline
    Collision system & \TwoS/\OneS  & \ThreeS/\OneS         \\ 
    \hline
    $pp, \sqs = 5.36\tev$ &  $0.256\pm0.002\pm0.008$ &  $0.115\pm0.001\pm0.008$  \\
    $p$O, \sqsnn = 9.62\tev &  $0.208\pm0.030\pm0.006$ &  $0.069\pm0.030\pm0.005$   \\
    OO, \sqsnn = 5.36\tev   &  $0.127\pm0.024\pm0.008$ &  $0.058\pm0.020\pm0.005$  \\
    \hline
  \end{tabular}\end{center}
\label{tab:Tab1}
\end{table}

\begin{table}[h]
    \centering
    \caption{Double ratios of the \ThreeS/\OneS and \TwoS/\OneS production cross sections in $p$O and OO collisions to $pp$ collisions, as shown in Fig.~\ref{fig:Fig4}. The first uncertainty is statistical and the second is systematic.}
    \begin{tabular}{c | c | c}
    \hline
        Collision system &
        $\left[\frac{\sigma_{\TwoS}}{\sigma_{\OneS}}\right]_{\mathrm{A}'\mathrm{A}} \Big/ \left[\frac{\sigma_{\TwoS}}{\sigma_{\OneS}}\right]_{pp}$ &
        $\left[\frac{\sigma_{\ThreeS}}{\sigma_{\OneS}}\right]_{\mathrm{A}'\mathrm{A}} \Big/ \left[\frac{\sigma_{\ThreeS}}{\sigma_{\OneS}}\right]_{pp}$ \\
        \hline
        $p$O & $0.81\pm0.12\pm0.04$ & $0.54\pm0.17\pm0.05$ \\
        OO   & $0.50\pm0.10\pm0.03$ & $0.45\pm0.15\pm0.05$ \\
    \hline
    \end{tabular}
    \label{tab:Tab2}
\end{table}

\clearpage


\addcontentsline{toc}{section}{References}
\bibliographystyle{LHCb/LHCb}
\bibliography{main,LHCb/standard,LHCb/LHCb-PAPER,LHCb/LHCb-CONF,LHCb/LHCb-DP,LHCb/LHCb-TDR}

\newpage
\centerline
{\large\bf LHCb collaboration}
\begin
{flushleft}
\small
R.~Aaij$^{39}$\lhcborcid{0000-0003-0533-1952},
M.~Abdelfatah$^{71}$,
A.S.W.~Abdelmotteleb$^{59}$\lhcborcid{0000-0001-7905-0542},
C.~Abellan~Beteta$^{53}$\lhcborcid{0009-0009-0869-6798},
F.~Abudin\'en$^{61}$\lhcborcid{0000-0002-6737-3528},
T.~Ackernley$^{63}$\lhcborcid{0000-0002-5951-3498},
A.A.~Adefisoye$^{71}$\lhcborcid{0000-0003-2448-1550},
B.~Adeva$^{49}$\lhcborcid{0000-0001-9756-3712},
M.~Adinolfi$^{57}$\lhcborcid{0000-0002-1326-1264},
P.~Adlarson$^{87,44}$\lhcborcid{0000-0001-6280-3851}
% ...

{\footnotesize \it

% ...
$^{39}$Nikhef National Institute for Subatomic Physics, Amsterdam, Netherlands\\
% ...
$^{44}$National Center for Nuclear Research (NCBJ), Warsaw, Poland\\
% ...
$^{49}$Instituto Galego de F{\'\i}sica de Altas Enerx{\'\i}as (IGFAE), Universidade de Santiago de Compostela, Santiago de Compostela, Spain\\
% ...
$^{53}$Physik-Institut, Universit{\"a}t Z{\"u}rich, Z{\"u}rich, Switzerland\\
% ...
$^{57}$H.H. Wills Physics Laboratory, University of Bristol, Bristol, United Kingdom\\
% ...
$^{59}$Department of Physics, University of Warwick, Coventry, United Kingdom\\
% ...
$^{61}$School of Physics and Astronomy, University of Edinburgh, Edinburgh, United Kingdom\\
$^{62}$School of Physics and Astronomy, University of Glasgow, Glasgow, United Kingdom\\
$^{63}$Oliver Lodge Laboratory, University of Liverpool, Liverpool, United Kingdom\\
% ...
$^{71}$Syracuse University, Syracuse, NY, United States\\
% ...
$^{87}$Department of Physics and Astronomy, Uppsala University, Uppsala, Sweden, associated to $^{62}$\\
% ...
}
\end{flushleft}

\end{document}